\documentclass{aa}  

\usepackage{graphicx}
\usepackage{txfonts,natbib}
\begin{document}

   \title{Laboratory rotational ground state transitions of  NH$_3$D$^+$ and CF$^+$}

   \subtitle{}

   \author{A. Stoffels
          \inst{1,2}
          \and L. Kluge
          \inst{1}
	\and S. Schlemmer
	\inst{1}
	\and S. Brünken
	\inst{1}
          }

   \institute{I. Physikalisches Institut, Universität zu Köln, Zülpicher Str. 77, 50937 Köln, Germany
              \email{bruenken@ph1.uni-koeln.de}
         \and
            Radboud University, Institute for Molecules and Materials, FELIX Laboratory, Toernooiveld 7, 6525 ED Nijmegen, The Netherlands\\
             }

   \date{Received ; accepted}

 
  \abstract
   {}
   {This paper reports accurate laboratory frequencies of the rotational ground state transitions of two astronomically relevant molecular ions, NH$_3$D$^+$ and CF$^+$.}
   {Spectra in the millimeter-wave band were recorded by the method of rotational state-selective attachment of He-atoms to the molecular ions stored and cooled in a cryogenic ion trap held at 4~K. The lowest rotational transition in the {\it A} state  (ortho state) of NH$_3$D$^+$ ($J_K=1_0-0_0$), and the two hyperfine components of the ground state transition of CF$^+$  ($J=1-0$) were measured with a relative precision better than 10$^{-7}$.}
   {For both target ions the experimental transition frequencies agree with recent observations of the same lines in different astronomical environments. In the case of NH$_3$D$^+$ the high-accuracy laboratory measurements lend support to its tentative identification in the interstellar medium. For CF$^+$ the experimentally determined hyperfine splitting confirms previous quantum-chemical calculations and the intrinsic spectroscopic nature of a double-peaked line profile observed in the $J=1-0$ transition towards the Horsehead PDR. }
   {}

   \keywords{ Astrochemistry; Line: identification; molecular data; Methods: laboratory: molecular; ISM: molecules    
             }

   \maketitle
%

\section{Introduction}

Molecular cations are important constituents of the interstellar medium (ISM). Exothermic and barrierless ion-molecule reactions are the key drivers of the chemistry in many interstellar environments such as the diffuse medium, photon-dominated regions (PDRs) and dense, cold molecular clouds. In many cases molecular cations have also proven to be excellent probes of the physical conditions in a variety of astronomical sources. Recent examples are the detection of ArH$^+$ in the diffuse galactic and extragalactic interstellar medium as a tracer of almost purely atomic hydrogen gas \citep{BSO2013,SNM2014,MMS2015}, and the observation and use of the {\it ortho/para}-ratio of H$_2$D$^+$ as a chemical clock to determine the age of a molecular cloud \citep{BSC2014}. One essential prerequisite for the detection and use of new molecular probes is the knowledge of highly accurate ($\Delta \nu / \nu< 10^{-6}$) rotational transition frequencies from  laboratory experiments. In the case of reactive molecular ions, standard absorption spectroscopy through electrical discharges is often hampered by too  low production yields of the target ions, and spectral contamination from a multitude of simultaneously generated molecular species. In particular rotational ground state transitions, often the most readily observed lines in the cold interstellar medium, are difficult to measure in the high excitation conditions of a laboratory plasma. These limitations can be overcome by employing sensitive action spectroscopic techniques on mass-selected ions in cryogenic ion traps. In our group we have in the past years developed several of these techniques based on laser induced reactions \citep[LIR;][]{SKL1999,SLR2002} and demonstrated their potential to determine highly accurate rotational transition frequencies \citep{ARM2008,GKK2010,GKK2013,JAW2014}. 

Here we present high-resolution measurements of the rotational ground state transitions of  NH$_3$D$^+$ and  CF$^+$, whose astrophysical importance as well as previous laboratory studies will be summarized in the remainder of this section. We then, in Section \ref{sec_lab},  introduce the experimental setup and the method of rotational state-selective attachment of helium atoms to mass-selected ions stored in a cryogenic (4~K) 22-pole ion trap instrument \citep{ABK2014}, previously applied in our group for rotational spectroscopy of C$_3$H$^+$ \citep{BKS2014}, and present experimental results and spectroscopic analyses on both ions. In Section \ref{discussion} we will discuss our results in the context of recent astrophysical observations.

\subsection{The ammonium ion, NH$_4^+$}
The ammonium ion, NH$_4^+$, is related to nitrogen chemistry in the interstellar medium, which has obtained renewed interest in the last years due to astronomical observations of light nitrogen hydrides (NH, NH$_2$, NH$_3$) with {\it Herschel}/HIFI in cold protostellar \citep{HMB2010,BCH2010} and diffuse gas \citep{PBC2010,PDM2012}. In the cold interstellar medium  NH$_4^+$, formed by subsequent hydrogenation reactions starting from N$^+$ with H$_2$, is assumed to be the gas-phase precursor of the ubiquitous ammonia molecule, NH$_3$, the dominant product of its dissociative recombination with electrons \citep[e.g.,][and references therein]{LeBourlot1991,MWM2012}. Ammonia, once formed either by the above gas-phase route or by sublimation from ices in warmer regions of the ISM, can be transformed back to NH$_4^+$ by exothermic proton transfer reactions, dominantly with H$_3^+$, so that the chemistry of these species is closely linked. The high deuterium fractionation of ammonia seen in many dark clouds \citep{RTC2000,SOO2000,LRG2002,TSM2002,LGR2006} is caused primarily by similar transfer reactions with deuterated forms of H$_3^+$, producing deuterated variants of NH$_4^+$, followed by dissociative recombination with electrons \citep{RC2001,FPW2006,SHC2015}. 
Whereas NH$_4^+$, a spherical top molecule, is not observable by radio astronomy, its mono- (and doubly-) deuterated forms have comparatively large permanent electric dipole moments \citep[0.26~D in the case of NH$_3$D$^+$,][]{NA1986} and possess rotational transition lines in the mm- and submm-wavelength range. The monodeuterated variant NH$_3$D$^+$ was recently tentatively detected toward the massive star-forming region Orion-IRc2 and the cold prestellar core B1-bS \citep{CTF2013}. In both sources a spectral feature at 262816.7~MHz was observed and identified as the $J_K=1_0-0_0$ rotational ground state transition of  NH$_3$D$^+$ based on accompanying high-resolution infrared data of the $\nu _4$ vibrational band \citep{DCH2013} that significantly reduced the uncertainty of the transition frequency prediction compared to an earlier study \citep{NA1986}.  Nevertheless, the reported frequency uncertainty of $\pm 6$~MHz is comparable to the typical spacing between spectral features in a line-rich source like Orion, allowing for ambiguity in the line assignments. Here we present a direct laboratory measurement of the observed rotational transition, with $2$ orders of magnitude higher accuracy, lending additional support to the identification of NH$_3$D$^+$ in space.

\subsection{The  fluoromethylidynium cation, CF$^+$}
The fluoromethylidynium cation, CF$^+$, has been detected in a variety of interstellar environments, namely in two photodissociation regions (PDRs) \citep{NSM2006,GPG2012}, in several diffuse and translucent galactic gas clouds \citep{LPG2014,LGP2015}, in the envelope of a high-mass protostar \citep{FBS2015}, and recently even in an extragalactic source \citep{MKB2016}. The main interest in CF$^+$ observations lies in its formation pathway: CF$^+$ is formed by an exchange of the fluorine atom of hydrogen fluoride (HF) in a bimolecular reaction with ionized atomic carbon (C$^+$), and thus can serve as a molecular tracer for HF and C$^+$ \citep{NWS2005,NSM2006,GPG2012,LGP2015}. Both of these species are keystones in astrochemistry, HF is the main reservoir of fluorine in the ISM and a useful probe of molecular H$_2$ column density, whereas C$^+$ is the main gas-phase reservoir of carbon in the diffuse interstellar medium, providing the dominant cooling process via its $^2P_{3/2}-^2P_{1/2}$ fine structure transition at 158~$\mu$m (1.9~THz). Both the C$^+$ fine structure line, and the rotational ground state line of HF (at around 1.2~THz) are not observable from the ground, whereas CF$^+$ (with a rotational constant of around 51.3~GHz) has many accessible transitions at mm-wavelengths. Its rotational spectrum is well studied up to 1.6~THz by absorption spectroscopy through magnetically enhanced negative glow discharges of CF$_4$ and H$_2$ \citep{PAH1986,CCP2010,CCB2012}. The CF$^+$ ground state transition, however, had not been observed in the laboratory plasma. In fact, the highest resolution  measurement of this line so far comes from its astronomical observation in the Horsehead PDR \citep{GPG2012}. Interestingly, the authors found a double-peaked feature at the expected CF$^+$ ($J=1-0$) transition frequency in this narrow-line source, which they originally attributed to kinematic effects based on similarity to the observed C$^+$ line profile. Shortly thereafter, however, quantum chemical calculations revealed that the line profile can be explained spectrocopically by hyperfine splitting (hfs) caused by the non-zero nuclear spin ($I=1/2$) of the fluorine nucleus \citep{GRG2012}. The low temperatures achievable in our cryogenic ion trap experiments enabled us to resolve the hyperfine splitting of the CF$^+$  ground state transition experimentally for the first time, and to compare the derived fluorine spin rotation constant to the quantum chemical calculations.


\section{Laboratory Measurements and Results}
\label{sec_lab}

The ground state rotational transitions of NH$_3$D$^+$ and CF$^+$ were recorded in a cryogenic 22-pole ion trap instrument \citep[FELion, almost identical to the one described in][]{ABK2014} with the method of rotational state-selective attachment of He atoms to the cold and mass-selected ions that was developed in our laboratory and recently applied to measure several rotational lines of linear C$_3$H$^+$ \citep{BKS2014}. In the following we will provide a brief description of the general method, before giving specific details on the NH$_3$D$^+$ and CF$^+$ experiments and analysis. 

Ions are produced in an ion storage source by electron impact ionization of suitable precursor gases. After selecting the target ions by a quadrupole mass-filter they enter the 22-pole ion trap in a short (few 10~ms) pulse, where they are stored and efficiently cooled close to the ambient temperature (4~K) by collisions with helium gas, which is constantly admitted to the trap. Ternary collision processes of the stored ions with helium lead to the formation of weakly bound ion-He complexes, which in turn can dissociate again by binary collisions with the helium atoms (collision induced dissociation, CID). At the low temperature and high number density of He ($\sim 5 \cdot 10^{14}$~cm$^{-3}$) in the trap, up to 10\% of the primary ions are converted to ion-He complexes at equilibrium, reached typically after several 100~ms storage time. Rotational spectra are recorded by irradiating the stored ions continuously with radiation from a narrow-bandwidth mm-wave source, and counting the number of He-ion complexes formed after a fixed storage time as a function of excitation frequency. For this we make use of the fact  that the helium attachment rate depends on the internal (rotational) excitation of the ion, as described in \cite{BKS2014}. To account for long-term fluctuations in ion signal we perform alternating measurements at the set frequency and a fixed offset frequency and give the spectroscopic signal in relative units scaled to the ion-He counts recorded at the offset position (depletion signal $S=\frac{N_{\mbox{off}}-N_{\mbox{on}}}{N_{\mbox{off}}}$).

\subsection{Rotational transitions of NH$_3$D$^+$}

\begin{figure}
   \centering
   \includegraphics[width=\hsize]{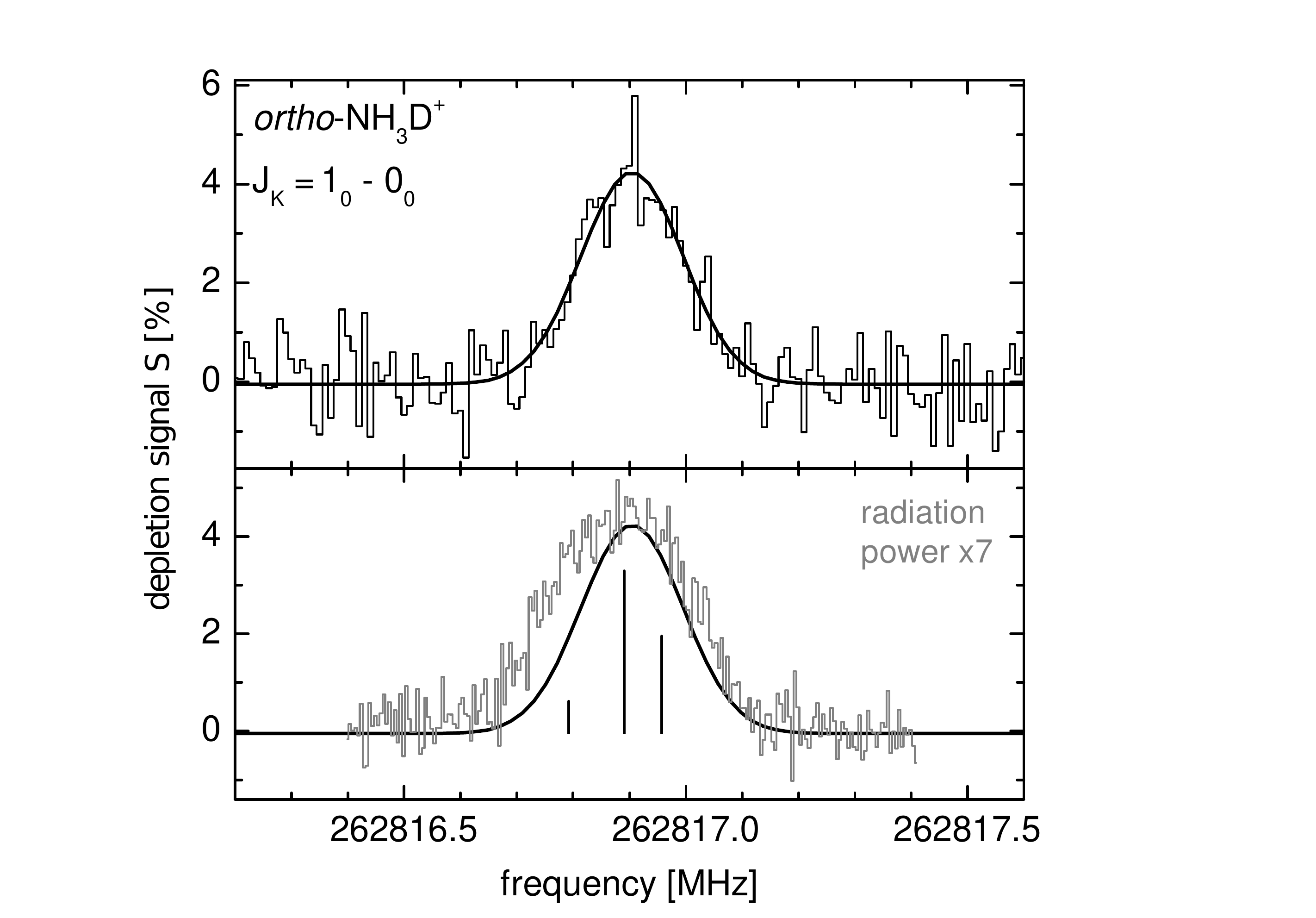}
      \caption{Top: Measured rotational ground state transition $J_K=1_0-0_0$ of ortho-NH$_3$D$^+$ (black histogram).  About 10000 mass-selected NH$_3$D$^+$ ions were stored for 600~ms in the cryogenic  ion trap (4~K) in the presence of a high number density of helium ($\sim 4-7\cdot10^{14}$~cm$^{-3}$) and continuous irradiation with tunable narrowband mm-wavelength radiation ($\sim 30$~$\mu$W power in the trap region). The absolute number of  helium complexes formed over the storage time was $\sim 400-700$. The spectrum is averaged over 100 iterations, and the line center position was extracted from fitting a Gaussian line-shape function to the data (black line). Bottom: The same line recorded with seven times higher radiation power (grey histogram). The line is broadened due to a low-frequency shoulder attributed to partial saturation of  individual hyperfine components. The estimated hfs components due to the deuterium nuclear quadrupole moment and their relative intensities are shown as black bars (C. Puzzarini, private communication). The Gaussian fit function of the unsaturated line is shown for comparison (black line). }
         \label{F_NH3D+line}
   \end{figure}

NH$_3$D$^+$ was produced in the ion storage source from a mixture of ammonia (NH$_3$) and D$_2$ at a ratio of 2:1 and pressure of $\sim 5 \cdot 10^{-6}$~mbar in combination with an electron impact ionization energy of 20~eV. For the measurements around 10000 mass-selected NH$_3$D$^+$ ions (m=19~u) were stored in the trap at 4~K  in the presence of helium gas at number densities between $(4-7)\cdot 10^{14}$~cm$^{-3}$, resulting in $400-700$ observed  NH$_3$D$^+ \bullet$He complexes after a storage time of 600~ms . Ions were irradiated continuously with the output of a narrowband ($\Delta \nu<<1$~kHz) frequency multiplier chain (Virginia Diodes Inc. WR9.0M-AMC in combination with a WR2.8x3 tripler, yielding a total frequency multiplication by 27) driven by a rubidium-clock referenced microwave synthesizer (Rohde und Schwarz SMF100A). For the spectrum shown in Fig. \ref{F_NH3D+line} the radiation was coupled to the trap via a conical horn antenna through a 0.8 mm thick CVD diamond window (diameter 18~mm) without additional focusing elements, resulting in $\sim30$~$\mu$W radiation power in the trap region. 

The rotational $J_K=1_0-0_0$  ground state transition of ortho-NH$_3$D$^+$  recorded in this way is shown in the upper panel of Fig. \ref{F_NH3D+line}. We observe a peak depletion signal of $\sim4$~\% in the number of  NH$_3$D$^+ \bullet$He complexes at the transition frequency of $262\,816.904(15)$~MHz, with a linewidth (FWHM) of 213(13)~kHz obtained by fitting a Gaussian line-shape function to the data. Assuming a purely Doppler broadened line this would correspond to a kinetic ion temperature of 24(3)~K, considerably higher than the ambient trap temperature (4~K, corresponding to a linewidth of 86~kHz). The larger linewidth can partly be explained by unresolved underlying hyperfine structure (hfs) due to the nuclear quadrupole moments of the N and D nuclei in NH$_3$D$^+$, with the main contribution from the D nuclei leading to a splitting of the $J_K=1_0-0_0$ transition into three main components (with 1:5:3 intensity ratio) separated by $\sim 100$~kHz (C. Puzzarini, private communication). We indeed observe the appearance of a low-frequency shoulder in the spectrum at the expected position of the weak $F=1-0$ hyperfine component when increasing the radiation power by a factor of 7, i. e. to a level where the individual  hfs components become saturated (see lower panel of Fig. \ref{F_NH3D+line}). The higher excitation power was achieved by using an elliptical focusing mirror (f=43.7~mm) in combination with a larger diameter (63~mm) z-cut Quartz window to couple the radiation into the trap region. We checked carefully for possible saturation-induced shifts in the derived transition frequency of the unresolved hfs-multiplett. At the lower power settings they are of the order of several kHz and are contained in the stated 15~kHz uncertainty of the fitted line measurement (Table \ref{nh3d+}). \\

The prolate symmetric rotor NH$_3$D$^+$ contains three identical hydrogen nuclei with spin $I_H=1/2$ and consequently occurs as an ortho and a para spin isomer with $I_{H,total}=3/2$ and 1/2, respectively. Rotational levels of A symmetry ($K=0$ and $K=3n$) belong to the ortho isomer, those of E symmetry ($K=3n\pm 1$) to the para isomer. Since collisional or radiative transitions between the two spin isomers are highly forbidden, and spin-conversion reactions are  likely hindered by substantial barriers and therefore do not proceed \citep{BD1987}, we expect that the cooled and stored NH$_3$D$^+$ ions retain their 1:1 formation ortho/para-ratio, i.e. that half of them belong to the para isomer. 
For the para $J_K=2_1-1_1$ ground state transition we expected a depletion signal  similar to that of the observed ortho ground state line. However,  integration down to an rms of 0.3~\% (using 15~kHz steps, and a nominal trap temperature T=4~K) in a 3~MHz scan around the predicted line position at $525\,572.0(3)$~MHz resulted in only a tentative $2\sigma$ detection of a weak feature at $525\,571.16(3)$~MHz, i.e. within $3\sigma$ of its predicted value. 
Possible reasons for the significantly lower depletion signal are that: i) the achievable radiation power at 525.6~GHz is about a factor 15 less than at 262.8~GHz, resulting in insufficient radiative pumping at the transition frequency; or ii) the change in ternary rate coefficient is significantly lower between higher $J$ (or $K$) levels . Similar arguments hold for the next higher ortho transition, $J_K=2_0-1_0$, which was not detected after integrating down to an rms-level of 0.9\% (at 15 kHz scan step size) in a 2.5~MHz bandwidth centered on the predicted $525\,589.65(13)$~MHz line position, even when increasing the nominal trap temperature to 9~K, i.e. increasing the thermal population of the initial $1_0$ level to 20~\%.\\

\begin{table*}
\caption{\label{nh3d+}Measured and predicted$^{a}$ rotational transitions of NH$_3$D$^+$ (in the vibrational ground state).}
\centering
\begin{tabular}{llllccc}
\hline\hline
$J$ & $K$ & $J'$ & $K'$  & Frequency [MHz] & o-c [kHz]$^{b}$ & reference \\
\hline
 1 & 0 & 0 & 0 & 262816.904(15) & 0 & exp. this work\\
 2 & 1 & 1 & 1 & 525571.97(28) &  & pred. this work \\
 2 & 0 & 1 & 0 & 525589.65(13) & & pred. this work\\
 3 & 1 & 2 & 1 & 788247.58(49) & & pred. this work\\
 3 & 0 & 2 & 0 & 788274.10(51) & & pred. this work\\
\hline
\end{tabular}
\tablefoot{Numbers in parentheses represent experimental or predicted $1\sigma$ uncertainties.$^{a}$ Predictions up to $E_{low}=30$~cm$^{-1}$.
$^{b}$ observed-calculated (o-c) differences with respect to the spectroscopic parameters given in Table \ref{nh3dpar}.}
\end{table*}

The measured $J_K=1_0-0_0$ transition frequency of NH$_3$D$^+$ (see Table \ref{nh3d+}) was analysed together with the previously reported IR ro-vibrational data from \cite{DCH2013} with the {\it spcat/spfit} program suite \citep{Pic1991} employing a prolate symmetric rotor Hamiltonian and neglecting hyperfine terms. The derived spectroscopic parameters of the vibrational ground state are given in Table \ref{nh3dpar} together with those obtained from analysing the ro-vibrational data alone. As expected, inclusion of the one high-resolution rotational transition significantly improves the accuracy of the $B_0$ rotational constant, whereas all higher order terms retain their uncertainty, and the purely $K$-dependent terms $A_0$ and $D_K$ needed to be kept fixed in the analysis.  Even though the inclusion of the tentatively detected para $J_K=2_1-1_1$ ground state transition does change the spectroscopic parameters and predicted transition frequencies within less than 3 times their predicted $1\sigma$ uncertainty, we did not include this line in the analysis because of its very low signal-to-noise value. Predictions of astronomically relevant rotational transition frequencies with lower state energies up to 30~cm$^{-1}$ are presented in Table \ref{nh3d+}.

\begin{table*}
\caption{\label{nh3dpar} Ground-state spectroscopic parameters of NH$_3$D$^+$.}
\centering
\begin{tabular}{lcccc}
\hline\hline
\multicolumn{2}{l}{Parameter}& This work & \citet{DCH2013} \\
\hline
$A_0$  &  (MHz)       & [175438.5464]$^{a}$   	&  [175438.5464]$^{a}$\\
$B_0$  & (MHz)         &  131412.1315(130) &  131412.08(88)           \\
$D _J$ & (MHz)	& 1.8397(53)	&	1.8396(112)			 \\
$D _{JK}$ & (MHz)	& 4.420(82)	&	4.404(88)			 \\
$D _{K}$ & (MHz)	& [0.0]$^{a}$	&	 [0.0]$^{a}$			 \\
\hline
wrms		&		&	0.94	& \\
\hline
\end{tabular}
\tablefoot{Numbers in parentheses represent $1\sigma$ uncertainties.
The ground state lines from this work (Table \ref{nh3d+}) were fitted together with ro-vibrational data from \citet{DCH2013}.
$^{a}$ Parameter was kept fixed in the fit to the value from \citet{NA1986}.}
\end{table*}

\subsection{The ground state rotational transition of CF$^+$}

CF$^+$ is efficiently produced in the ion storage source from pure tetrafluoromethane (CF$_4$ synthetic grade, at a pressure of $\sim 2 \cdot 10^{-5}$~mbar) at an ionization energy of 40~eV. About 8000 mass-selected CF$^+$ ions  (m=31~u) were stored in the 22-pole ion trap at a nominal temperature of 4.2~K in the presence of helium at a number density of $4\cdot 10^{14}$~cm$^{-3}$. Under these conditions around 400 CF$^+\bullet$He complexes are formed after 600~ms storage time. We used the output of a solid-state frequency tripler (Virginia Diodes Inc. AMC 375) pumped with an atomic-clock referenced synthesizer (Rohde und Schwarz SMF100A) to record the spectrum of the CF$^+$ $J=1-0$  line in a 2~MHz window (scanned with 10~kHz steps) centered at $102\,587.53$~MHz, i.e. the line position predicted from the earlier mm-wave measurements of higher $J$ transitions \citep{CCB2012}.\\

 \begin{figure}
   \centering
   \includegraphics[width=\hsize]{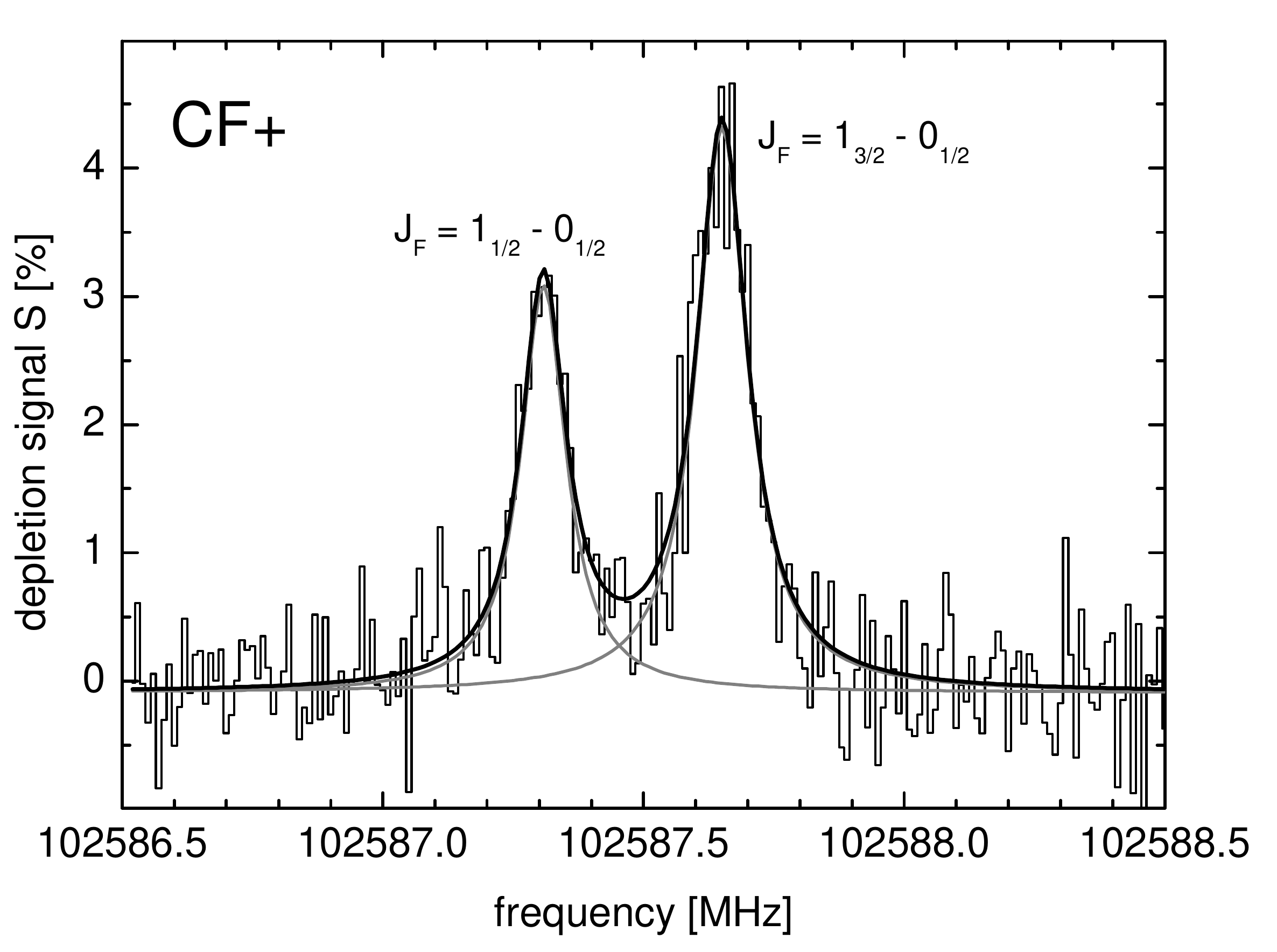}
      \caption{Rotational ground state transition $J=1-0$ of CF$^+$ with resolved hyperfine structure (black histogram) measured by storing about 8000 mass-selected CF$^+$ ions for 600~ms in a cryogenic 22-pole ion trap (4.2~K) in the presence of a high number density of helium ($\sim4\cdot10^{14}$~cm$^{-3}$) and continuous irradiation with tunable narrowband mm-wavelength radiation ($\sim 10$~$\mu$W power in the trap region). The relative depletion signal for both lines is of the order of a few percent, with an absolute number of around 400 helium complexes formed over the storage time. The spectrum is averaged over 320 iterations, and the line center positions were extracted from fitting a Voigt line shape function to the data (grey curves).}
         \label{F_CF+line}
   \end{figure}

The resulting spectrum of the CF$^+$ ground state transition is shown in Fig. \ref{F_CF+line}. The two hyperfine components attributed to the $J_F=1_{1/2}-0_{1/2}$ and $J_F=1_{3/2}-0_{1/2}$ ($\mathbf{F=J+I}$, with $I=1/2$ the fluorine nuclear spin) are clearly resolved. The line center frequencies of the two components given in Table \ref{cf+} were obtained from fitting a double Voigt line shape function to the data. The Voigt line shape is a consequence of combined line broadening due to the ions kinetic temperature (a Doppler temperature T=36(15)~K can be derived from the Gaussian contribution to the linewidth, $\sigma = 34(8)$~kHz, but this value is likely overestimated because of  incomplete deconvolution of the Gaussian and Lorentzian line parameters), and due to power broadening, which is of the same order of magnitude ($\gamma=38(10)$~kHz for the stronger hfs component) at the radiation power P$\sim$10~$\mu$W used to record the lines shown in Fig. \ref{F_CF+line}. In fact, when increasing the power P we observe the expected $\sqrt{P}$-dependent increase of the linewidths (with the Lorentz linewidth (HWHM) given by $\gamma=\mu E/ \hbar$, $\mu$ the transition moment of the transition, and $E\propto\sqrt{P}$ the electric field strength of the exciting radiation), with the linewidth of the $J_F=1_{3/2}-0_{1/2}$ line twice as large as that of the $J_F=1_{1/2}-0_{1/2}$ component, as expected from the twice as large transition moment.  This saturation effect can also explain why we do not observe the theoretically expected 1:2 intensity ratio for the two hfs components shown in Fig. \ref{F_CF+line}. At P$>$250$~\mu$W the two hyperfine components can no longer be resolved.  \\

\begin{table*}
\caption{\label{cf+}Measured and predicted rotational transitions of CF$^+$.}
\centering
\begin{tabular}{llllccc}
\hline\hline
$J$ & $J'$ & $F$ & $F'$ & Frequency [MHz] & o-c [kHz]$^{a}$ & reference\\
\hline
1 & 0 & 1/2 & 1/2  & 102587.308(4) & 0.6 & exp. this work\\
1 & 0 & 3/2 & 1/2  & 102587.649(3)& 0.7 & exp. this work\\
2 & 1 &       &         & 205170.520(20) & 9 & \citet{CCP2010}\\
3 & 2 &       &         & 307744.351(20) & -20 & \citet{CCP2010}\\
4 & 3 &       &         & 410304.565(20) & 9 & \citet{CCP2010}\\
5 & 4 &       &         &  512846.493(20)& -16 & \citet{CCB2012}\\
6 & 5 &       &         & 615365.671(9) &  & pred. this work \\
7 & 6 &       &         &  717857.486(9)&  & pred. this work \\
8 & 7 &       &         &  820317.395(11)&  & pred. this work \\
9 & 8 &       &         &  922740.843(13)&  & pred. this work \\
10 & 9 &       &         & 1025123.275(30) & 1 & \citet{CCB2012}\\
11 & 10 &       &         & 1127460.137(30) & 8 & \citet{CCB2012}\\
12 & 11 &       &         &  1229746.851(27)&  & pred. this work \\
13 & 12 &       &         &  1331978.890(35)&  & pred. this work \\
14 & 13 &       &         &  1434151.691(45)&  & pred. this work \\
15 & 14 &       &         & 1536260.652(100) & 24 & \citet{CCB2012}\\
\hline
\end{tabular}
\tablefoot{Numbers in parentheses represent experimental or predicted $1\sigma$ uncertainties.\\
$^{a}$ observed-calculated (o-c) differences with respect to the spectroscopic parameters given in Table. \ref{cfpar}.}
\end{table*}

We performed a least-squares fit of the experimental rotational transition frequencies of CF$^+$ from this work and  all previous studies (compiled in Table \ref{cf+}) to a diatomic rotor Hamiltonian including the fluorine spin-rotation interaction constant $C_I$. Our best fit rotational constant $B_0$ and centrifugal distortion constant $D_0$ (Table \ref{cfpar}) agree within their $1\sigma$ uncertainties with those from \cite{CCB2012}, and as in this study the sextic centrifugal distortion constant $H_0$ was kept fixed to that of \cite{GPS1986} derived from ro-vibrational measurements. The spin-rotation constant $C_I$ is experimentally determined to an accuracy of 3~kHz, reflecting the high accuracy of our experimental line positions, and is in perfect agreement with the quantum-chemically calculated value reported in \cite{GRG2012}, also given in Table \ref{cfpar}. From the spectroscopic constants all experimental transition frequencies can be reproduced to within their experimental uncertainty (Table~\ref{cf+}). The Table also gives rest frequency predictions for additional rotational transitions of CF$^+$ up to 1.5~THz (not including hfs splitting).
   
\begin{table*}
\caption{\label{cfpar} Ground-state spectroscopic parameters of CF$^+$.}
\centering
\begin{tabular}{lccccc}
\hline\hline
\multicolumn{2}{l}{Parameter}& This work & \citet{CCB2012} & \citet{GRG2012}\\
\hline
$B_0$  &  (MHz)       & 51294.14713(93)     &  51294.14672(98) &\\
$D_0$  & (kHz)         & 189.9277(54) &            189.9267(47) &   \\
$H_0$ & (Hz)    	& [0.07]$^{a}$ &   [0.07]$^{a}$  & \\
$C_I$ & (kHz)	& 227.3(33)	&				& 229.2$^{b}$ \\
\hline
wrms  &			& 0.50 & & \\
\hline
\end{tabular}
\tablefoot{Numbers in parentheses represent $1\sigma$ uncertainties.
The two hfs components from this work were fitted together with 7 rotational lines from \citet{CCP2010} and \citet{CCB2012} (see Table \ref{cf+}). 
$^{a}$ Fixed to the experimental value by \citet{GPS1986}. $^{b}$ Calculated value at the CCSD(T)/cc-pCV6Z level.}
\end{table*}

\section{Discussion and Conclusions}
\label{discussion}

In this work we present a highly accurate ($\Delta \nu / \nu =6\cdot 10^{-8}$) experimental frequency for the $J_K=1_0-0_0$ rotational ground state transition of ortho-NH$_3$D$^+$ that can directly be compared to available astronomical observations. The narrow spectral emission feature at 262816.73(10)~MHz detected by \cite{CTF2013} toward the cold prestellar core B1-bS agrees within twice its experimental uncertainty with our experimental value of $262\,816.904(15)$~MHz. The small discrepancy might be attributed to uncertainty in the LSR velocity of the source. Using the laboratory rest frequency we obtain a v$_{\mbox{LSR}}$ of 6.7~km/s from the astronomical line. This value is well in the range of observed v$_{\mbox{LSR}}$ in single dish observations of Barnard B1-b \citep[e.g.,][]{LGR2006,DGR2013,CBA2014}, but falls somewhat between the systemic velocities of 6.3~km/s and 7.2~km/s associated with the two embedded B1-bS and B1-bN cores derived from high-spatial resolution observations, of e.g. N$_2$H$^+$, by \cite{HH2013}. Nevertheless, from a purely spectroscopic view our laboratory measurements support the assignment of the astronomical lines observed in B1-bS and Orion IRc2 to NH$_3$D$^+$. Ancillary support for this identification can be obtained by the observation of additional rotational transitions, for which accurate rest frequencies can now be predicted from our spectroscopic analysis (Tables \ref{nh3d+} and \ref{nh3dpar}). The  para ground state transition is here of particular interest, since it will give information on the ortho/para-ratio of NH$_3$D$^+$, which is intimately linked to the spin state ratios of ammonia and H$_3^+$ and their deuterated variants \citep{SHC2015}. Unfortunately, both the $J_K=2_1-1_1$ para ground state line and the next higher ortho transition $J_K=2_0-1_0$, both at around 525.6~GHz, are extremely difficult to observe from ground. \\

Both transitions were covered in the HEXOS line survey towards Orion KL \citep{CBN2014,CBN2015} using the HIFI instrument \citep{GHP2010} on the Herschel Space Observatory \citep{PRP2010}. Spectra extracted from the publicly available archival data show no spectral features with significant intensity above the rms value of $\sigma=20$~mK (in $T_A^*$ units) at the two predicted transition frequencies. This is in accordance with the expected estimated peak intensity of around 25~mK for this line  based on the intensity of the $J_K=1_0-0_0$ line ($\sim200$~mK) observed by \cite{CTF2013} with the 30m IRAM telescope towards the IRc2 position (only $3\arcsec$ offset of the HEXOS survey pointing position). Our estimate assumes that NH$_3$D$^+$ is present in the compact ridge region (with a source size of a few arcseconds), based on the observed v$_{\mbox{LSR}}=9$~km/s and narrow-linewidth usually attributed to this component. In this case the $J=2-1$ emission lines seen with the $40\arcsec$ Herschel beam are severely more diluted than the $J=1-0$ line observed with the IRAM 30m telescope ($10\arcsec$ beam). We furthermore assumed thermalisation of the rotational levels with an excitation temperature of >80~K (except for the ortho:para ratio, which was set to 1) because of the rather small Einstein coefficients for spontaneous emission $A_{ij}$ of the order $4\cdot 10^{-5}$~s$^{-1}$ and an  H$_2$ density of $10^{6}$~cm$^{-3}$ in the compact ridge \citep{CBN2014}. The cold prestellar cloud B1-bS, exhibiting much fewer and narrower lines at sub-millimeter wavelengths than the complex Orion region, is likely a better target to search for confirming transitions of NH$_3$D$^+$ (the two 525.6~GHz lines should be of similar intensity as the detected ortho ground state line assuming a rotational excitation temperature of 12~K). Their detection, however, needs to wait for the next generation receivers onboard the airborne SOFIA observatory. An alternative is to search for higher deuterated variants of the ammonium ion, i.e. NH$_2$D$_2^+$ and ND$_3$H$^+$, as suggested by \cite{CTF2013}. The feasibility of spectroscopic studies on these species, the prerequisite for an astronomical search, will be discussed below. \\

Owing to the low ion temperatures and thereby narrow Doppler widths achievable in our trap experiments, we were able to resolve the two hyperfine components of the CF$^+$ $J=1-0$ rotational ground state transition, and extract their transition frequencies to within $3-4$~kHz, i.e. to a relative accuracy better than $10^{-7}$.  From this we accurately determined the spin-rotation interaction constant to $C_I=227(3)$~kHz. This value is in excellent agreement with a high-level theoretical value of 229.2~kHz that was used by \cite{GRG2012} to account for an observed double-peaked line structure in the CF$^+$ $J=1-0$ transition towards the Horsehead PDR \citep{GPG2012}. Thus our spectroscopic experiments confirm the intrinsic nature of the astronomically observed line structure, ruling out that they stem from kinematic effects.\\

State-selective attachment of He atoms to mass-selected, cold molecular ions, used in this work for the spectroscopy of NH$_3$D$^+$ and CF$^+$, is a very powerful and general method for rotational spectroscopy. Due to the low internal ion temperature it is particularly well suited to probe rotational ground state transitions, which are difficult to observe with standard absorption experiments through high-temperature plasmas. 
In contrast to the more established action spectroscopic method of laser induced reactions (LIR) \citep{SLR2002}, the method directly probes purely rotational transitions, which for LIR is only possible in specific cases \citep[as for H$_2$D$^+$,][]{ARM2008} or through IR-mm-wavelength double resonance techniques \citep{GKK2013,JAW2014}. Moreover, for LIR a suitable endothermic reaction scheme is needed, whereas the attachment of He to the cold ions, and, even more importantly, a rotational state dependency of the attachment process needed for the spectroscopic scheme, has been observed in our laboratory for a considerable number of cations. These are, apart from NH$_3$D$^+$ and CF$^+$ presented here, C$_3$H$^+$ \citep{BKS2014}, CO$^+$, HCO$^+$, CD$^+$ \citep{PhDKluge}, CH$_2$D$^+$, and CD$_2$H$^+$ (discussed in forthcoming publications). The observable depletion signal for a given rotational transition depends on various experimental parameters and intrinsic properties of the studied molecular ion, as we have thoroughly investigated on known rotational transitions of CD$^+$ and HCO$^+$ \citep{PhDKluge}. Under the premise that the participating rotational states do exhibit a difference in the ternary attachment rate, the depletion signal will be the stronger, the more the population ratio of the two levels changes, upon resonant excitation, from the initial thermal Boltzmann distribution (i.e. without radiative excitation). In the best case, when radiative processes dominate over collisional ones, the population ratio of the two states reaches ``radiative'' equilibrium given by the ratio of their statistical weights. The method therefore profits from the low achievable ion temperature, and is best suited for ground state transitions with large rotational energy spacing.  \\

Although the presented method is very sensitive in the sense that only a few thousand mass-selected molecular ions are needed to record rotational spectra, searching for unknown transitions over ranges larger than a few 10 MHz is very time-consuming owing to the long experimental cycle times (around 2~s per frequency point) and the multiple iterations needed to achieve a sufficient signal-to-noise ratio for detecting depletion signals of the order of typically only several percent. A very promising approach towards recording rotational transitions of new species, like the higher deuterated variants of NH$_4^+$, is to significantly reduce the experimental search range by obtaining accurate predictions for pure rotational transitions from preceding measurements of their ro-vibrational spectra with high accuracy via LIR (if applicable) or LIICG \citep[Laser Induced Inhibition of Complex Growth, as described for ro-vibrational spectroscopy in][]{ABK2014,AYB2015,JKS2016}. That frequency accuracies below 1~MHz can be achieved in this way has been demonstrated in our group on the examples of CH$_2$D$^+$ \citep{GKK2010,GKK2013}, deuterated variants of H$_3^+$ \citep{JKS2016}, and by determining the ground-state combination differences of CH$_5^+$ \citep{AYB2015} by using a narrow-linewidth OPO as infrared radiation source, in the latter two studies referenced to a frequency comb \citep{AKS2012}, and taking advantage of the low ion temperatures in the trap instruments. Once accurate predictions of the rotational transitions are available, they can be recorded directly with the method of state-selective attachment of He atoms presented here, or by LIR IR-mm-wavelength double resonance methods if applicable, thus providing highest accuracy and resolution needed for a comparison to radio-astronomical observations, as demonstrated here in the case of NH$_3$D$^+$ and CF$^+$. \\

\begin{acknowledgements}
     We thank H.S.P Müller for providing his analysis of the NH$_3$D$^+$ ro-vibrational data, C. Puzzarini and J. Gauss for data on the NH$_3$D$^+$ hyperfine splitting parameters, and P. Schilke for extracting the Herschel/HIFI archival spectra. {\it Herschel} is an ESA space observatory with science instruments provided by European-led Principal Investigator consortia and with important participation from NASA. S.B. and S.S. acknowledge financial support from the Deut\-sche For\-schungs\-ge\-mein\-schaft, DFG, via the priority program SPP 1573, and L.K. from the DFG collaborative research grant SFB 956 (sub-project B2). For the construction of the FELion instrument, the authors gratefully acknowledge the work done over the last years by the electrical and mechanical workshops of the institute workshops in Cologne as well as a collaboration with J. Oomens and B. Redlich (FELIX laboratory, Radboud University, Nijmegen).
\end{acknowledgements}


\bibliographystyle{aa}
\bibliography{astro,mm}

\clearpage

\end{document}